\newif\ifINCLUDE
\INCLUDEtrue

\newif\ifINCLUDENOT
\INCLUDENOTfalse

\newif\ifPAPERBACK
\PAPERBACKtrue 

\newif\ifINCLUDESPARETEXT 
\INCLUDESPARETEXTfalse

\newif\ifKINDLE
\newif\ifNOTKINDLE

\KINDLEfalse \NOTKINDLEtrue

\ifKINDLE
\fi
\ifNOTKINDLE
\fi

\newif\ifkeypoint
\keypointfalse

\newif\ifboxexplainer
\boxexplainertrue

\newif\ifseparate
\separatefalse

\newif\ifusecolourfigures
\usecolourfigurestrue

\newif\ifcolourinformation
\colourinformationtrue

\newif\ifCOMMENTS
\COMMENTSfalse

\newif\ifSHOWNOTES
\SHOWNOTESfalse

\newif\ifSHOWPVALUES
\SHOWPVALUESfalse

\newif\ifincludechnoisychannel
\includechnoisychannelfalse

\newif\ifincludechnoiseFREEchannel
\includechnoiseFREEchannelfalse

\newif\ifEXERCISES
\EXERCISESfalse

\newif\ifSHOWEDITS
\SHOWEDITSfalse

\newif\ifSHOWEDITP
\SHOWEDITPfalse

\newif\ifSHOWDELETES
\SHOWDELETESfalse

\newif\ifSHOWDELETEP
\SHOWDELETEPfalse

\newif\ifSHOWJVSDELETES
\SHOWJVSDELETESfalse

\newif\ifSHOWJVSEDITS
\SHOWJVSEDITSfalse

\newif\ifSHOWQUERIES
\SHOWQUERIESfalse

\newif\ifSHOWDELETESCO
\SHOWDELETESCOtrue

\newif\ifSHOWEDITSCO
\SHOWEDITSCOtrue

\newif\ifSHOWQUERIESCO
\SHOWQUERIESCOtrue

\newif\ifSHOWALLCO
\SHOWALLCOtrue
\SHOWALLCOfalse

\ifSHOWALLCO
\SHOWQUERIESCOtrue
\SHOWEDITSCOtrue
\SHOWDELETESCOtrue
\else
\SHOWQUERIESCOfalse
\SHOWEDITSCOfalse
\SHOWDELETESCOfalse
\fi

\newif\ifFIG
\newif\ifSYNOPSIS

\newif\ifTheoremsSummary
\TheoremsSummaryfalse

\newif\ifTTD
\TTDfalse
\ifTTD
\listoffigures
\fi

\documentclass[a4paper,11pt]{article} 
\usepackage {fullpage}

\usepackage{cancel} 

\usepackage{setspace}

\usepackage{geometry} 
	\FIGfalse 
	\SYNOPSISfalse

	\usepackage{latexsym}
	\usepackage{amsfonts}
	\usepackage{amsmath}
	\usepackage{subfig} 
	\usepackage{makeidx}
\usepackage{url} 

\usepackage[super,sort]{natbib}

\usepackage{ulem} 
\usepackage[pdftex]{graphicx}
\usepackage[small,compact]{titlesec} 
 \usepackage{fancyhdr}

\usepackage{subfig}
\usepackage{leading}
\usepackage{amsmath}
\usepackage{amsfonts}
\usepackage{boxedminipage} 
\usepackage{makeidx}

\usepackage{float}

\usepackage[usenames,dvipsnames,svgnames,table]{xcolor}

\usepackage{relsize} 

\usepackage{color}

\usepackage{gensymb} 

\usepackage{lipsum}

\usepackage{amsthm}
\usepackage{thmtools}

\usepackage{setspace}

    \usepackage{enumitem}

\newlength{\RoundedBoxWidth}
\newsavebox{\GrayRoundedBox}

   {\setlength{\RoundedBoxWidth}{\dimexpr#1}
    \begin{lrbox}{\GrayRoundedBox}
       \begin{minipage}{\RoundedBoxWidth}}%
   {   \end{minipage}
    \end{lrbox}
    \begin{center}
    \begin{tikzpicture}%
       \draw node[draw=black,fill=black!5,rounded corners,%
             inner sep=2ex,text width=\RoundedBoxWidth]%
             {\usebox{\GrayRoundedBox}};
    \end{tikzpicture}
    \end{center}}
    
   {\setlength{\RoundedBoxWidth}{\dimexpr#1}
    \begin{lrbox}{\GrayRoundedBox}
       \begin{minipage}{\RoundedBoxWidth}}%
   {   \end{minipage}
    \end{lrbox}
    \begin{center}
    \begin{tikzpicture}%
       \draw node[draw=black,fill=black!5,rounded corners,%
             inner sep=2ex,text width=\RoundedBoxWidth]%
             {\usebox{\GrayRoundedBox}};
    \end{tikzpicture}
    \end{center}}
    
\usepackage{amssymb,amsmath}

\usepackage[EULERGREEK]{sansmath}

\ifCOMMENTS

\else

\fi

\ifCOMMENTS

\else

\fi

\ifSHOWJVSEDITS 

\else

\fi

\ifSHOWJVSDELETES 
\usepackage{soul}
\newcommand{\JVSdelete}[1]{{\color{red}\st{#1}}}
\else
\newcommand{\JVSdelete}[1]{\unskip}
\fi

\ifSHOWEDITS

\else

\fi

\ifSHOWEDITP

\else

\fi

\ifSHOWDELETES
\usepackage{soul}
\newcommand{\delete}[1]{{\color{red}\st{#1}}}
\else
\newcommand{\delete}[1]{\unskip}
\fi

\ifSHOWDELETEP
\usepackage{soul}
\newcommand{\deletep}[1]{{\color{blue}\st{#1}}}
\else
\newcommand{\deletep}[1]{\unskip}
\fi

\ifSHOWQUERIES
\newcommand{\CO}[1]{{\color{red}[CO: #1]}}
\newcommand{\query}[1]{{\color{blue}[CO: #1]}}
\else
\newcommand{\CO}[1]{\unskip}
\newcommand{\query}[1]{\unskip}
\fi
\ifSHOWQUERIESCO
\newcommand{\queryCO}[1]{{\color{purple}[CO: #1]}}
\else
\newcommand{\queryCO}[1]{}
\fi

\ifSHOWDELETESCO
\usepackage{soul}
\newcommand{\deleteCO}[1]{{\color{red}\st{#1}}}
\else
\newcommand{\deleteCO}[1]{}
\fi

\ifSHOWEDITSCO
\newcommand{\editCO}[1]{{\color{teal}#1}}
\else
\newcommand{\editCO}[1]{{#1}}
\fi
\newif\ifFIG
\FIGtrue

\makeindex

\renewcommand{\ni}{\noindent}

\newcommand{\ie}{{i.e.\:}}

\newcommand{\etal}{{et al.\ }}

\newcommand{\var}{{\rm var}}

\usepackage{xspace}

\leading{14pt}

\let\tempone\description
\let\temptwo\enddescription
\renewenvironment{description}{\tempone\addtolength{\itemsep}{-4.0pt}}{\temptwo}

\newcommand{\be}        {\begin{equation}  }
\newcommand{\ee}        {\end{equation}	}
\newcommand{\bea}       {\begin{eqnarray}  } \newcommand{\eea}       {\end{eqnarray}    }
\newcommand{\beal}       {\begin{align}} 
\newcommand{\eeal}       {\end{align}}

\newcommand{\jsubsection}[1]{\vspace{0.05in}\ni  {\bf #1}.}

\renewcommand{\belowcaptionskip}{-15pt}

\newcommand{\captionfonts}{\small}
\makeatletter  
\long\def\@makecaption#1#2{%
  \vskip\abovecaptionskip
  \sbox\@tempboxa{{\captionfonts #1 #2}}%
  \ifdim \wd\@tempboxa >\hsize
    {\captionfonts #1 #2\par}
  \else
    \hbox to\hsize{\hfil\box\@tempboxa\hfil}%
  \fi
  \vskip\belowcaptionskip}
\makeatother   

\newcommand{\bw}{      {\bf w} }

\renewcommand{\tt}{{\theta}}

\newcommand{\nn}	{\nonumber}

\newcommand{\E}{{\rm E}}

\newcommand{\ybar}{\overline{y}}

\renewcommand{\ni}{\noindent}

\renewcommand{\em}{\it} 

\AtBeginDocument{}
\makeatletter
\DeclareOldFontCommand{\rm}{\normalfont\rmfamily}{\mathrm}
\DeclareOldFontCommand{\sf}{\normalfont\sffamily}{\mathsf}
\DeclareOldFontCommand{\tt}{\normalfont\ttfamily}{\mathtt}
\DeclareOldFontCommand{\bf}{\normalfont\bfseries}{\mathbf}
\DeclareOldFontCommand{\it}{\normalfont\itshape}{\mathit}
\DeclareOldFontCommand{\sl}{\normalfont\slshape}{\@nomath\sl}
\DeclareOldFontCommand{\sc}{\normalfont\scshape}{\@nomath\sc}
\makeatother

\usepackage{setspace,lipsum}
\usepackage{graphicx}
\begin{document}

\thispagestyle{empty}

\title{
{\bf  
Methods for 
Estimating Neural Information
}
}

\author{
James V Stone, 
University of Sheffield, England.\\
j.v.stone@sheffield.ac.uk\\
File: main\_NeuralInfoTheory\_v47Methods2022.tex
}

\date{}
\maketitle




\thickmuskip=1mu plus 1mu minus 1mu 
\thinmuskip=1mu plus 1mu minus 1mu 


\onehalfspacing

\newcommand{\BOOKCOVERpath}{/Users/JimStone/Documents/BOOKS/ALLBookCoversFRONT/Small}

\newcommand{\bookwidth}{0.4}
\newcommand{\bookgap}{\hspace{0.1in}}

\newcommand{\bookVgap}{\vspace{0.05in}}

\begin{abstract}
\noindent
Estimating the Shannon information associated with individual neurons is a non-trivial problem. Three key methods used to estimate the mutual information between neuron inputs and outputs are described, and a list of further readings is provided.
\end{abstract}

\section{Neural Information Methods}

\markboth{Neural Information Methods \vspace{-0.2in}
}{Neural Information Methods} 
\index{information estimation}%

Consider  a temporal sequence of stimulus values $x$\deleteCO{,} and the resultant neuron outputs $y$\editCO{,} 
which can be either a sequence of continuous values or a sequence of spikes. 
The total Shannon entropy $H(y)$ in the outputs is essentially  a global measure of how much the response sequence varies over time. In contrast, the noise entropy $H(y|x)$  is a measure of how much variation in the response sequence remains after the stimulus value $x$ at each point in time has been taken into account. Therefore, the difference between $H(y)$  and $H(y|x)$ is  the amount of variation in the response sequence that can be attributed to the stimulus sequence. 
This difference is the mutual information\cite{stonearXivinfotutorial2018,StoneInformationBook2022} between $x$ and $y$, 
\bea
	I(x,y) & = & H(y) - H(y|x) )  \text{ bits},  \label{eqMI803a}
\eea
where all logarithms are base 2, which ensures that information is measured in bits; one bit provides enough information to choose between two equally probable alternatives. 

In practice, it will prove useful to know that mutual information can be obtained from  two other equations. 
Somewhat counter-intuitively, $I(x,y)$ is also given by 
 the difference between  $H(x)$ (the entropy of the stimulus values)  and $H(x|y)$  
(the entropy in the stimulus values $x$ that remains after the responses $y$ have been taken into account),
\bea
	I(x,y)	& = & H(x) - H(x|y) )  \text{ bits}. \label{eqMI803ab}   \label{eqMI803abcx}
\eea
Finally, it can be shown that 
\bea
	I(x,y)	& \leq & 0.5 \, \log \, (1+ SNR)  \text{ bits},  \label{eqMI803abc}
\eea
where SNR is the signal\editCO{-}to\editCO{-}noise ratio (see Section \ref{secupper}), with equality  if each variable is independent and has a  Gaussian distribution. 

\newpage
\noindent
The mutual information can be estimated using 
 three broad strategies\cite{borst1999information},  
which provide:
\begin{enumerate}
\item  a direct estimate using Equation \ref{eqMI803a}, 
\item a lower bound using Equation \ref{eqMI803ab},
\item an upper bound using Equation \ref{eqMI803abc}. 
\end{enumerate}
For simplicity, stimulus values 
are represented as $x$ here,
so that $y \, = \, g(x)+\eta$,
where $g$ is a neuron transfer function\deleteCO{,} 
and $\eta$ is a noise term. 
\ifFIG
\begin{figure}[b!] 
\begin{center}
\includegraphics[width=0.8\textwidth] {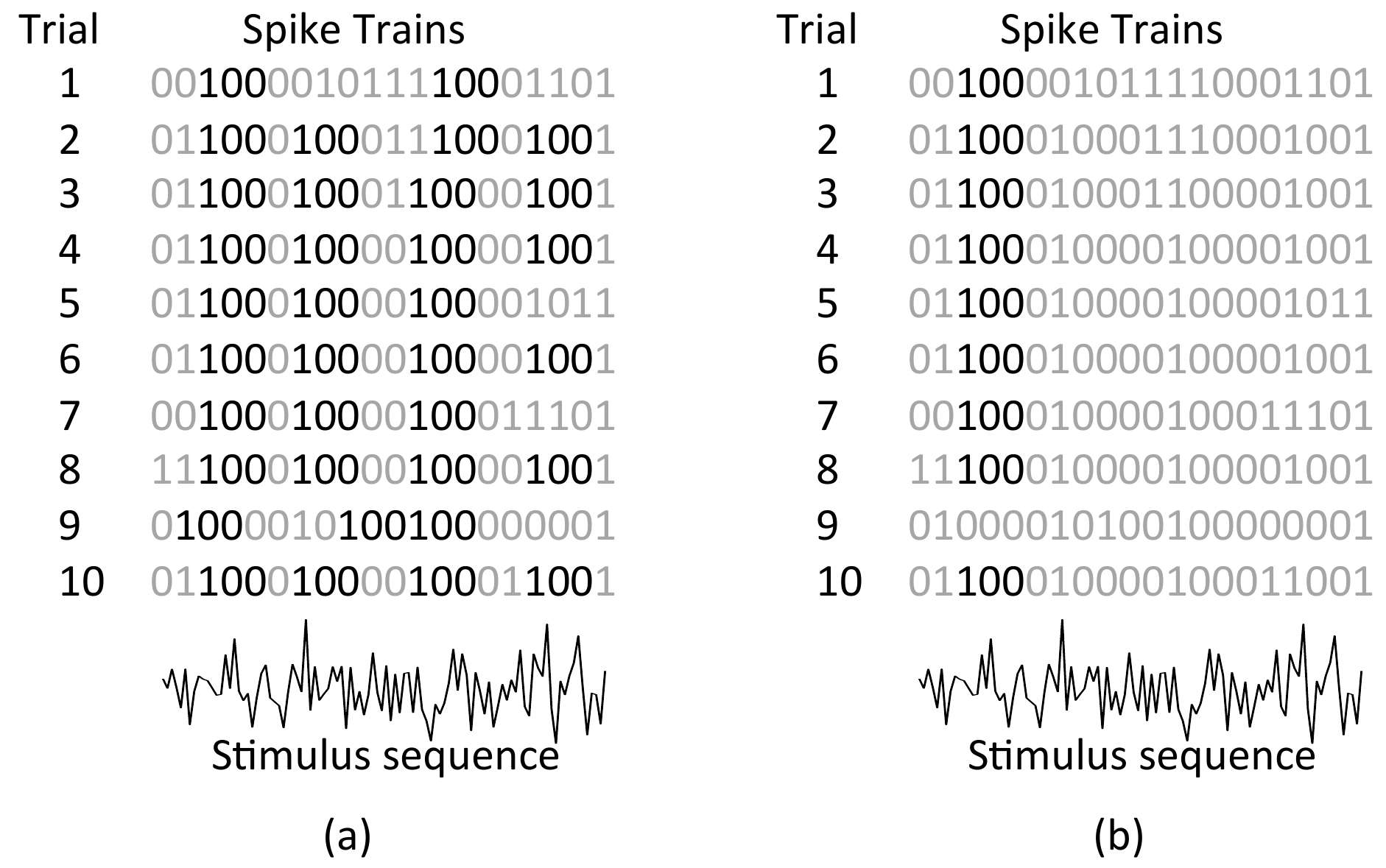} 
\caption{
The direct method (schematic). The same stimulus sequence is repeated for $N=10$ trials\deleteCO{,} and the $N$ response sequences are recorded\deleteCO{, where}\editCO{;} a spike is represented as 1 and no spike as 0.  \\
\editCO{(}a) Total entropy $H(y)$ is estimated from the probability of particular spike trains within a long unique spike train sequence (which is the concatenation of 10 trials here). The probability $p(y)$ of a particular $T$-element spike train $y$ is estimated as the number of instances of  $y$\deleteCO{,} expressed as a proportion of all $T$-element spike trains. For example, in the data above, there are 170 places where a \deleteCO{3}\editCO{three}-element spike train could occur, and there are 35 instances of the spike sequence $y=[100]$ (marked in bold), so $p(y)=35/170 \approx 0.206$.
\newline
\editCO{(}b) Noise entropy $H(y|x)$ is estimated from the conditional probability  of particular  spike trains.  The same stimulus value occurs at the same time in each of $N=10$ trials. Therefore, the conditional probability $p(y|x)$ of the response $y$ to a stimulus sub\deleteCO{-}sequence $x$ which starts at time $t$ is the number $N_{y}$ of trials which contain $y$ at time $t$\deleteCO{,} expressed as a proportion of the number $N$ of spike trains that begin at time $t$ (i.e. $p(y|x)=p(y|t)$).
For example, there are $N_{y}=9$ instances of  the spike sequence $y=[100]$ at $t=3$ (marked in bold),
so the conditional probability is $p(y=[100]|t=3) = 9/10=0.9$.
}
\label{figstrong1998DirectMethod1} 
\end{center}
\end{figure}
\fi

\section{The Direct Method}
\index{direct method}


\ifFIG
\begin{figure}[b!] 
\begin{center}
\includegraphics[width=0.6\textwidth] {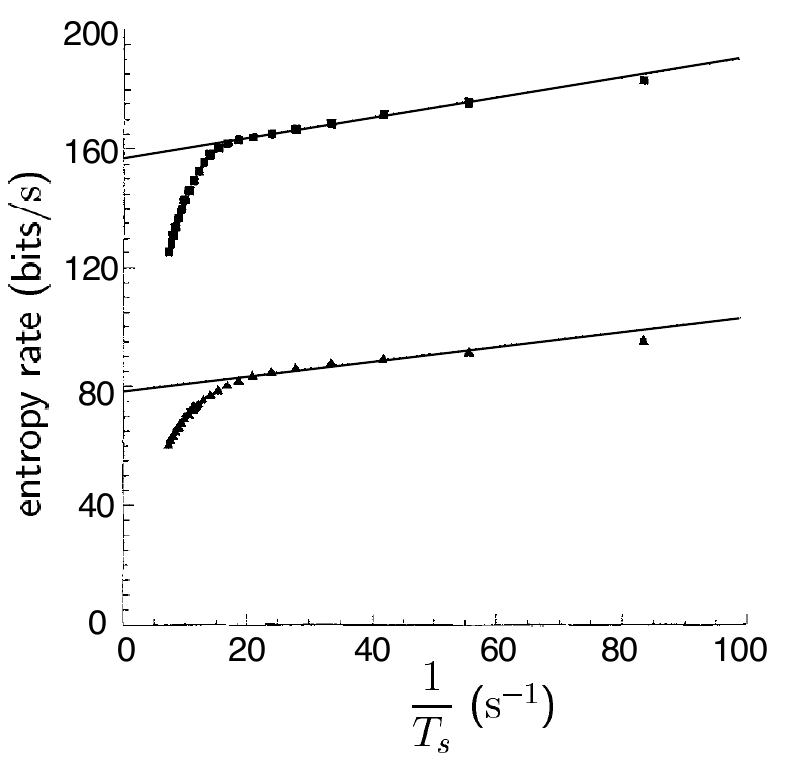} 
\caption{
The direct method. 
Entropy and noise entropy rates for a  visual neuron (H1 in the fly), responding to a randomly moving visual image. The filled circles in the upper trace show the full spike-train entropy rate for different values of $1/T$  (with $\Delta t = 3$\,ms). The straight line is an extrapolation to $1/T = 0$ (\ie $T \rightarrow \infty$)\deleteCO{,} and yields $H(y)$. 
The lower trace shows the spike\editCO{-}train noise entropy rate for different values of $1/T$, and the straight line is again an extrapolation to $1/T = 0$\deleteCO{,} and yields $H(y|x)$. 
The difference between the ordinate intercepts of the two straight lines is $H(y)-H(y|x)$\deleteCO{,} and is therefore the mutual information rate (Equation \ref{eqMI803a}).  Reproduced with permission from Strong \etal (1998)\cite{Strong1998}.
}
\label{figstrong1998DirectMethod2} 
\end{center}
\end{figure}
\fi

\vspace{0.1in}
\noindent
{\bf Estimating the Entropy of a Spike Train}. 
In physics, the entropy of a jar of gas is proportional to the volume of the jar. By analogy, we can treat a spike train as if it were a one\editCO{-}dimensional jar, so that spike train entropy is proportional to the amount of time $T$ over which the spike train is measured\editCO{:} $H(T,\Delta t) \, \propto \,  T$, where $\Delta t$ defines the temporal resolution used to measure spikes. 
Dividing $H(T,\Delta t)$ by $T$ yields the {\em entropy rate}, which converges to the entropy $H(y)$  for large values of $T$\editCO{;} specifically, 
\index{entropy rate}%
\bea
	   H(y) & = &  \lim_{T \rightarrow \infty} \frac{H(T,\Delta t)}{T}  \:\: \:\: \text{ bits/s}. \label{eqhconvergeA}
\eea 
 Strong \etal (1998)\cite{Strong1998} use arguments from statistical mechanics to show that 
a graph of $H(T,\Delta t)/T$ versus $1/T$ should yield a straight line (\deleteCO{also }see \editCO{also} Appendix A.8 in Bialek, 2012\cite{BialekBiophysicsSearchingPrinciples2012}).  
The $x$-intercept of this line is at $1/T \,=\, 0$, corresponding to a $y$-intercept\queryCO{set x and y as variables (italic)} of $H(T,\Delta t)/T$ at $T\,=\,\infty$, which is therefore  the entropy $H(y)$. 

The direct method usually involves two types of output sequences: {\em unique} and {\em repeated}. 
The unique spike train is a response to a long sequence of inputs; this is used to estimate the total spike train entropy. 
The repeated spike train sequence consists of spike trains obtained in response to $N$ repeats of a stimulus sequence; these are used to estimate the entropy of the noise in the spike train. However, if the repeated sequence is sufficiently long then the set of $N$ response sequences can be treated as a unique spike train, as in Figure \ref{figstrong1998DirectMethod1}.

\vspace{0.1in}
\noindent
{\bf Estimating Total Entropy $H(y)$}. 
The entropy  $H(T,\Delta t)$ for one value of $T$ is estimated from the probability $p(y^{i})$ of the $m_{T}$ different  observed  sequences $y^{1},\dots,y^{m_{T}}$ of length $T$\editCO{:} 
\bea
	H(T,\Delta t) & = &   \sum_{i=1}^{m_{T}} p(y^{i}) \log \frac{1}{p(y^{i})},
\eea
where $ p(y^{i})$ is the number of instances of the sequence $y^{i}$, expressed as a proportion of the number of
different sequences of length $T$ observed anywhere in the unique output sequence (see Figure \ref{figstrong1998DirectMethod1}a).

The entropy of the output sequence is found by estimating $H(T,\Delta t)/T$ for successively larger values of $T$\deleteCO{,} and  then extrapolating to find the entropy at $1/T\,=\,0$ (i.e. at $T\,=\,\infty$). In the limit $T \rightarrow \infty$,
\bea
	H(y) 
	& = &  \lim_{T \rightarrow \infty} \frac{H(T,\Delta t)}{T} \\
	\nn \\
	& = &  \lim_{T \rightarrow \infty}   \frac{1}{T} \sum_{i=1}^{m_{T}} p(y^{i}) \log \frac{1}{p(y^{i})}, \label{eqHy}
\eea
as shown by the upper line in Figure \ref{figstrong1998DirectMethod2}. 

\vspace{0.1in}
\noindent
{\bf Estimating Noise Entropy $H(y|x)$}. 
The stimulus  sequence $x$ is repeated $N$ times, so there are a total of $N$ similar response sequences.
The conditional (i.e. noise) entropy is estimated as 
\bea
	H(y|x) & \approx &\E_{t}[  H(y|x^{t}) ],
\eea
where $x^{t}$ is the stimulus sub\deleteCO{-}sequence starting at time $t$\deleteCO{,} and $y$ is the corresponding response.
Note that this average is taken over successive time indices between $t\,=\,1$ and $t\,=\,n-T$. 
$H(y|x^{t})$ is the entropy of the output sequences $y^{i}$ given $x^{t}$ (analogous to Equation \ref{eqHy})\editCO{:} 
\bea
	H(y|x^{t}) & = &  \lim_{T \rightarrow \infty}  \frac{1}{T} \sum_{i=1}^{m_{t}} p(y^{i}|x^{t})  \log \frac{1}{p(y^{i}|x^{t})}, \label{eqygx}
\eea 
where $p(y^{i}|x^{t})$ is the number of instances of the sequence $y^{i}$\deleteCO{,} expressed as a proportion of the number of
different sequences of length $T$ observed at time $t$ in the output sequences (see Figure \ref{figstrong1998DirectMethod1}b).
Note that the same stimulus value occurs at the same time in each trial, so\deleteCO{ that} $p(y|x^{t})\,=\,p(y|t)$.
As above, $H(y|x^{t})$ is found by evaluating the right\editCO{-}hand side of Equation \ref{eqygx}
 for successively larger values of $T$\deleteCO{,} and  extrapolating to find the entropy at $1/T\,=\,0$ (i.e. at $T\,=\,\infty$), 
 as shown by the lower line in Figure \ref{figstrong1998DirectMethod2}. 
Finally, mutual information is estimated from Equation \ref{eqMI803a}. 
\deleteCO{Also }\editCO{S}ee \editCO{also} Nemenman, Shafee, and Bialek (2002)\cite{NSBBialek2002}. 

\jsubsection{Assumptions} Inputs are repeated many times. Data are spike trains. 
\queryCO{need a noun here; please check suggestion and change if necessary JVS OK}\editCO{The estimation process m}akes no assumptions regarding the distribution of variables\deleteCO{,} and therefore requires large amounts of data.

\section{The Upper Bound Method} \label{secupper}
\index{upper bound method}%
If the noise $\eta$ in the output $y$ has an independent Gaussian distribution then the mutual information between $x$ and $y$ is maximised provided $x$ also has an independent Gaussian distribution. 
Thus, if the input $x$ is Gaussian and independent then the estimated mutual information provides an upper bound.
 Additionally, if each variable is Gaussian (but not necessarily independent) with a bandwidth of $W$\,Hz then its entropy is the sum of \editCO{the} entropies of its Fourier components\cite{FourierBookStone2021}.  

In common with the direct method, input sequences need to be  repeated many times, but the number $N$ of trials (repeats) required \editCO{here} is fewer\deleteCO{ here}. This is because a Gaussian distribution is defined in terms of its mean and variance, so, in effect,  we only need to estimate a few means and variances from the data. 

\subsection*{Estimating Output Signal Power}
\begin{enumerate}
\item Find the average output sequence $\ybar \,=\,   {1}/{N} \sum_{i=1}^{N} y^{i}$. 

\item Obtain\queryCO{Want to add `the' but it would take it over one line} Fourier coefficient  ($a(f),b(f)$) of $\ybar$ at each frequency $f$.
\item Estimate the power of each frequency $f$ as ${\mathcal S}({f}) \,=\,a({f})^{2}+b({f})^{2}$.
\end{enumerate}

\subsection*{Estimating Output  Noise Power}
\begin{enumerate}
\item Estimate \editCO{the} noise  $\eta^{i} \,=\, y^{i} - \ybar$ in each of the $N$ output sequences. 

\item Find \editCO{the} Fourier coefficient  ($a({f}),b({f})$) of $\eta^{i}$ at each frequency $f$. 
\item Estimate the power at each frequency $f$ as ${\mathcal N}^{i}({f})\,=\,a({f})^{2}+b({f})^{2}$.
\item Find the average power of each Fourier component 
\bea
	{\mathcal N}({f}) & = & \frac{1}{N} \sum_{i=1}^{N} {\mathcal N}^{i}({f}).
\eea
\end{enumerate}
Assuming a Nyquist sampling rate of $2W$\,Hz, estimate the mutual information $I(x,y)$ by summing over frequencies
\bea 
	R_{info} & = &  \sum_{f=0}^{W} \log  \left( 1+ \frac{{\mathcal S}({f})}  {{\mathcal N}({f})} \right)  \: \text{ bits/s}, \label{eqIxydecomp685tA}
\eea 
where $R_{info}$ $\geq$ $I(x,y)$, with equality if each variable is iid Gaussian. 
\index{Fourier analysis}%

\jsubsection{Assumptions} The response sequences to each of $N$ repeats of the same stimulus sequence are continuous. Each output sequence is Gaussian, but not necessarily independent (iid). 

\section{The Lower Bound Method}
\index{lower bound method}%
Unlike previous methods, this method does not rely on repeated presentations of the same stimulus, and \editCO{it} can be used for spiking or continuous outputs.  
In both cases, we can use the neuron inputs $x$ and outputs $y$ to estimate a linear decoding filter $\bw_{d}$. 
When the output sequence is convolved with this filter, it provides an estimate $x_{est} \,=\, \bw_{d}  \otimes y$ 
 of the stimulus $x$, 
where $\otimes$ is the convolution operator.   
We assume that $x \,=\, x_{est} + \xi_{est}$, 
so that the estimated noise in the estimated stimulus sequence is $\xi_{est} \,=\,x-x_{est}$. 

Assuming a bandwidth of $W$\,Hz and that values are transmitted at the Nyquist rate of $2W$\,Hz, we \deleteCO{proceed by }Fourier transform\cite{FourierBookStone2021} the stimulus sequence $x$ to find the signal power 
${\mathcal X}(f)$ at each frequency $f$\deleteCO{,} and Fourier transform $\xi_{est}$ to find the  power in the estimated noise  ${\mathcal M}(f)$  at each frequency. 
The mutual information is estimated by summing over frequencies\editCO{:}
\bea
	R_{min} & =  & H(x) - H(\xi_{est}) \label{eqHxxy} \\ 
				& = & 
				\sum_{f}   \log {\mathcal X}(f) -  
				\sum_{f}  \log {\mathcal M}(f)\\
				& = & 
				\sum_{f=0}^{W}  \log \frac{{\mathcal X}(f) }{{\mathcal M}(f) } \: \: \text{ bits/s},
				\label{Ixyasdfads}
\eea
where $R_{min} \,\leq\, I(x,y)$, with equality if each variable is iid Gaussian. 


%
%

\jsubsection{Assumptions} 
\editCO{The s}timulus sequence $x$ is Gaussian, but not necessarily independent (iid).  
Outputs are spiking or continuous.

\vspace{0.1in}
\noindent
{\bf Further Reading}. 
This is an extract from Principles of Neural Information Theory (2018)\cite{StoneNeuralInformationBook2018}, and is based on Strong \etal (1998)\cite{Strong1998}, Rieke \etal (1997)\cite{BialekBook1996}, 
Borst and Theunissen (1999)\cite{borst1999information}, Dayan and Abbot (2001)\cite{DayanAndAbbott2001}\editCO{,} and Niven \etal (2007)\cite{LaughlinEnergy2007}.
Relevant developments can be found in
Nemenman, Shafee, and Bialek (2002)\cite{NSBBialek2002}, 
Juusola \etal (2003, 2016)\cite{juusola2003rate,juusola2016electrophysiological}, 
 Ince \etal (2009)\cite{ince2009python}, Goldberg \etal (2009)\cite{goldberg2009spike},  Crumiller \etal (2013)\cite{crumiller2013measurement},  
Valiant and Valiant (2013)\cite{valiant2013estimating}, 
and  Dettner \etal (2016)\cite{timingcodespikeNature2016}. 
A tutorial account of information theory can be found on arxiv\cite{stonearXivinfotutorial2018}, and in these books\cite{StoneNeuralInformationBook2018,StoneInformationBook2022}.


\singlespace

\newcommand{\aaa}{0.25}
\begin{figure}[h!] 
\begin{center}
\subfloat{\includegraphics[  height = \aaa \textheight ] {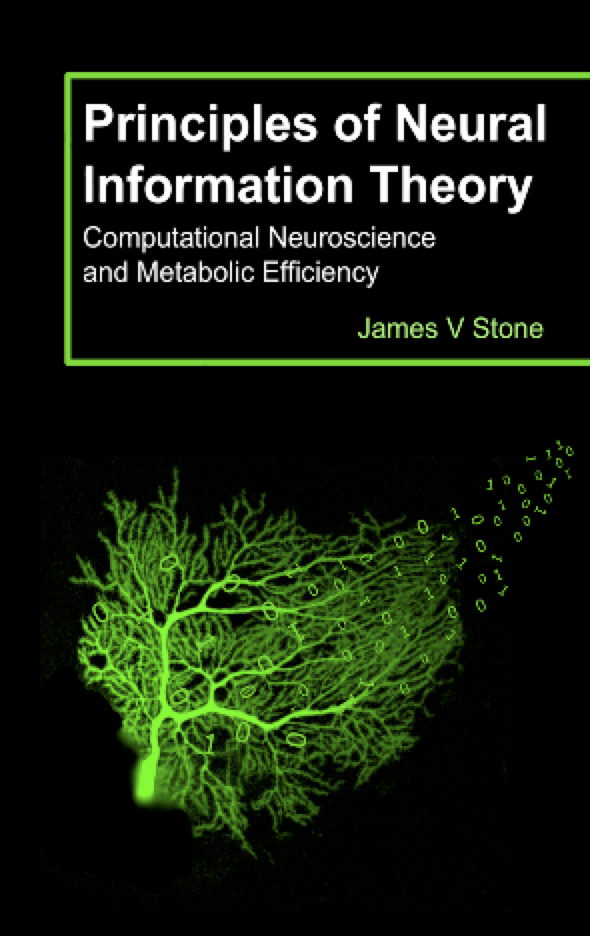} }
\subfloat{\includegraphics[ height = \aaa \textheight ] {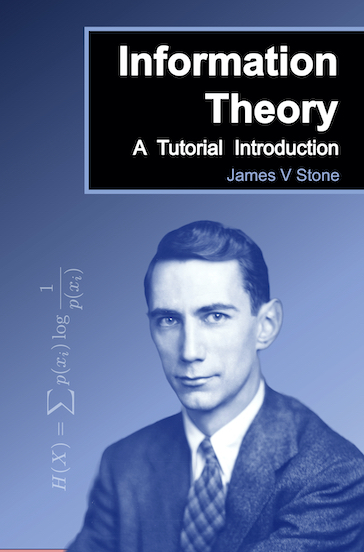} }
\subfloat{\includegraphics[ height = \aaa \textheight ] {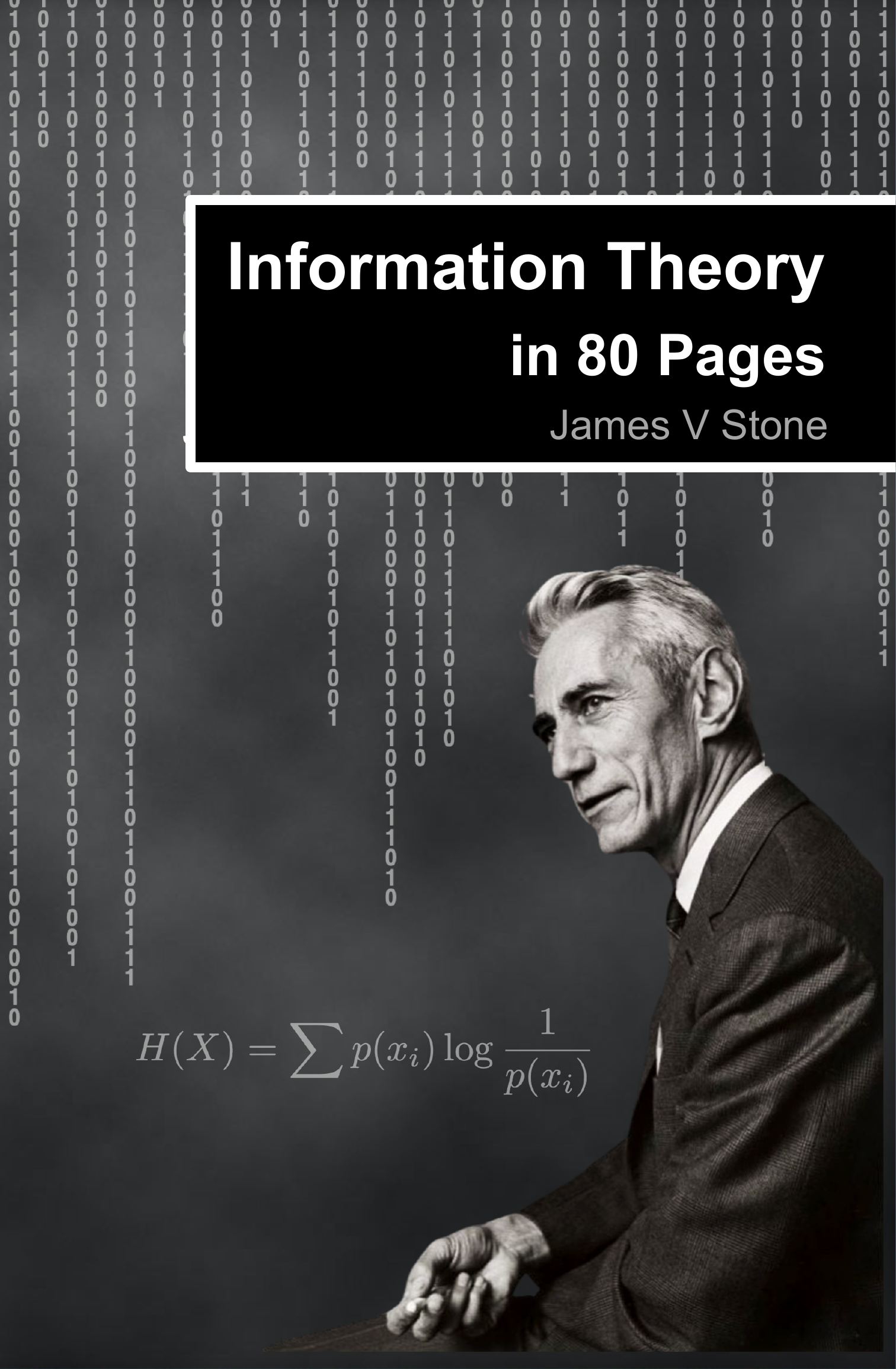} } 
\end{center}
\end{figure}


\end{document}